\documentstyle[preprint,aps]{revtex}
\def\prb{Phys. Rev. B}

\def\ebp{empirical bowing parameter}
\def\be{\begin{equation}}
\def\ee{\end{equation}}

\begin{document}
\draft
\preprint{CIEA-96/5 MC}
\title{
Electronic structure of the valence band of the II--VI wide band gap 
binary/ternary alloy interfaces}
\author{D. Olgu\'{\i}n and R. Baquero}
\address{
Departamento de F\'{\i}sica,\\
Centro de Investigaci\'on y de Estudios Avanzados del IPN,\\
A. P. 14-740, 07000 M\'exico D.F.
}
\maketitle

\begin{abstract}
We present an electronic structure calculation
of the valence band for some II--VI binary/ternary alloy
interfaces.
We use the empirical tight--binding method and
the surface Green's function matching method. For the ternary alloys
we use our previously set Hamiltonians 
they describe well the band gap change with composition
obtained experimentally. At the interface domain, we find 
three non--dispersive and two interface states besides
the known bulk bands. The non--dispersive states are reminiscent 
of the ones already obtained experimentally as well as theoretically, 
in (001)--oriented surfaces. We make use of the available theoretical
calculations for the (001)--oriented surfaces of the binary 
compounds and for the binary/binary interfaces
to compare our new results with.
\end{abstract}
 

\narrowtext

\section{Introduction}

In recent years, new semiconductor heterostructures have attracted
considerable interest. Multiple quantum well structures and superlattices
of II--VI compounds are the subject of intensive study because of their 
interesting optical properties \cite{tersoff,flores,brasil,ichino}. With 
these structures, energy gaps ranging from the UV to IR are accessible
\cite{brasil,ichino,pelhos}. In these systems the binary interfaces are
usually lattice mismatched. This lattice mismatch modifies the band 
alignments, and hence modifies the device optical properties.
In searching for the desired material parameters such as 
band gap, lattice matching to substrates, dielectric contact, carrier
mobility, etc., a large number of materials, have been investigated.
Recently, high--quality cubic--structured ternary and quaternary alloys
have been proposed as appropriate materials for heterostructures \cite{brasil,%
ichino}. Ternary alloys 
allow a certain control of the induced strain at the interface.

The deep understanding of the physics of 
the interface is important for the detailed study
of thermal, optical, and other properties of quantum--wells 
and superlattices. 
The electronic properties at solid--solid interfaces
depend sometimes even on details of the interaction between the two 
atomic layers from the different materials in contact. Our work can
be used as a starting point to analyze those details. These are
responsible for the characteristics of interface reconstruction, 
thermodynamic properties, degree of intermixing, stress, compound
formation, etc.

In previous work, we have studied the electronic structure of
the valence
band for the (001)--surface of several II--VI wide band gap
semiconductors \cite{prb50,prb51},
and different binary heterostructures \cite{infcs}.
We have obtained the (001)--projected electronic structure for both,
surfaces and binary interfaces using the known 
Surface Green's Function Matching (SGFM) method \cite{g-m}.

In the (001)--surfaces in addition to the well known bulk bands and surface 
resonances, we have described three 
different structures in the valence band region,  the so--called
surface induced bulk states ($B_h,\ B_l$, and 
$B_s$). We have shown that these states
owe their origin to the creation of the surface, that is, 
they depend on the surface through the boundary condition
(the wave function has to be zero at the surface),
but they are not surface resonances.
They are {\it surface--induced bulk states} \cite{prb50,prb51}.
Later we found that this kind of 
induced states appear at the interface domain as well. 
Therefore, more generally, we found that any frontier can induce these 
states. For that reason we have redefined them
as frontier--induced semi-infinite medium (FISIM) states
since they are not, strictly speaking, {\it bulk} (infinite medium) states.
These FISIM states do not show dispersion as a function of the wave
vector {\bf k} for the surfaces studied. This is theoretically and
experimentally shown for the (001)--oriented 
CdTe surface \cite{prb50,prb51,niles,gawlik}. 
For the binary/binary (001)--oriented II--VI compound
interfaces, in contrast, they show some 
clear dispersion \cite{infcs}. 

The interest of the present
work is twofold. Firstly, we want to make practical use of our recently
set found tight--binding Hamiltonians for the ternary alloys.  
They reproduce the known experimental change
with the composition of the band gap and they can be further used in
detailed studies of different physical problems as, for example,
the dependence of the transport properties on composition in 
quantum well structures that avoid stress. 
The second interest of 
this work is the study of the evolution of the FISIM states
from a non--dispersive character to a dispersive one as 
stress and different crystal composition enters into play. 
We show that, if we select a ternary alloy to produce
little stress and change only slightly the composition, the 
FISIM states do exists on both sides of the interface 
but do not show as much dispersion. So the existence of 
the FISIM states is due to the existence of a frontier alone
and the amount of dispersion is related to the existing stress
at the interface and on the chemical character of the 
interface partner. 

We will present in this work the valence band of some II--VI
(001)--binary/ternary alloy interfaces and we will concentrate 
in particular in the
FISIM states. The method used is discussed in our previous 
work. Here we only summarize the relevant features of it in
Section II for completeness; Section III is devoted to discuss 
our results. Finally, we give our conclusions in section IV.

\section{The Method}

To describe the interface between two semiconductor compounds, we make 
use of tight--binding Hamiltonians. The Green's function matching
method takes into account the perturbation caused by the surface or
interface exactly,
at least in principle, and we can use the bulk
tight--binding parameters (TBP) \cite{rafa,noguera,quintanar}.
This does not mean that we are
using the same TBP for the surface, or for the interface and the bulk. 
Their difference is taken into account through the matching of the 
Green's functions. We use the method in the form cast by 
Garc\'\i a--Moliner and Velasco \cite{g-m}. They make use of the transfer
matrix approach first introduced by Falicov and Yndurain \cite{falicov}.
This approach became very useful due to the quickly converging 
algorithms of L\'opez--Sancho {\it et al.} \cite{sancho} Following the
suggestions
of these authors, the algorithms for all transfer matrices needed to 
deal with these systems can be found in a straightforward way
\cite{trieste}.

The Green's function for the interface, $G_I$, is given by \cite{g-m},
\begin{equation}
\label{infcs}
G_I^{-1}=G_{s(A)}^{-1} + G_{s(B)}^{-1} - I_BH^iI_A - I_AH^iI_B,
\end{equation}
where 
$G_{s(A)}$ and $G_{s(B)}$ are 
the surface Green's function of medium A and B, respectively.
$-{\cal I}_AH^i{\cal I}_B$
and $-{\cal I}_BH^i{\cal I}_A$ are the Hamiltonian matrices 
that describe the interaction between the two media. In our model 
these are $20\times20$ matrices, the input TBP for these matrices 
are the average for those of the two media.
This is a reasonable approximation when both sides of the interface
have the same crystallographic structure and we take the same basis of 
wave functions. 

The tight--binding Hamiltonians for the 
II--VI ternary alloys
are described in detail in Ref. [18]. Briefly speaking, 
we have used the tight--binding 
method and, under certain conditions, the virtual crystal 
approximation to study the ternary alloys.
We have included an \ebp\ in the $s-$on site TBP of 
the substituted ion. This procedure gave us the correct behaviour of the 
band gap value with composition \cite{ternario}.
More exactly for the 
TBP of the ternary alloy, we take
\be
\overline E_{\alpha,\alpha'}(x)= x E_{\alpha,\alpha'}^{(1)}
+ (1-x)E^{(2)}_{\alpha,\alpha'}, \qquad \alpha,\ \alpha'=s,\ p^3,\ s^*
\label{vca}
\ee
for all but the $s-$on site TBP of the substituted 
ion. In eq. (\ref{vca}) $E^{(1,2)}_{\alpha,\alpha'}$ are the TBP
for the compound 1 (2); $\alpha,\ \alpha'$ are the atomic 
orbitals used in the basis set.

For the $s-$on site TBP of the substituted ion we use the 
following expression 
\be
\overline E_{s,\nu}(x,b_\nu)= \overline E_{s,\nu}(x) +
x(1-x)b_\nu, \qquad \nu=a,\ c
\ee
where $\overline E_{s,\nu}(x)$ is given by eq. (\ref{vca}) and
$b_\nu$ is the \ebp\ per each different substitution (anion--substitution $(a)$ 
or cation--one $(c)$). In Table \ref{ebp} we have the \ebp s used 
in this work. We do not introduce any 
further parameter \cite{ternario}. 

From the knowledge of the Green's function, the local
density of states can be calculated from its imaginary part integrating
over the two--dimensional first Brillouin zone, 
the dispersion relations can be obtained from the poles of the 
real part. We have
applied previously this formalism to surfaces \cite{prb50,prb51,rafa,noguera}, 
interfaces \cite{infcs,quintanar,rafa-prb} and 
superlattices \cite{rafa-moliner}.
Now we present our results.

\section{Results and discussion}

This section is devoted to the discussion of the interface--valence band of the
(001)--projected electronic band structure of II--VI binary/ternary 
alloy interfaces.
We will present in this paper the (001)--CdTe/CdSe$_{.15}$Te$_{.85}$, 
(001)--CdTe/Zn$_{.17}$Cd$_{.83}$Te,
(001)--ZnSe/ZnSe$_{.87}$Te$_{.13}$, and (001)--ZnSe/Zn$_{.85}$Cd$_{.15}$Se
interfaces in detail. The
interfaces studied have been chosen with a composition 
$(x)$ as to give a minimum stress. 
For the lattice parameter value of the materials considered see 
Table II. As we can see the induced stress is small, 
about 1\%. This magnitude of the induced stress allow us to 
ignore its effect in our 
calculation. 
The real bulk bands
as well as the FISIM states, should lie very closely to our 
calculated ideal case. 
We adopt the same convention for the interface domain as in Ref. 
\cite{infcs}.
That is to say, we consider nearest neighbors interactions in our bulk
Hamiltonians and, as a consequence, four atomic layers as the interface
domain, two belonging to medium A and two to medium B. To distinguish
between the different atomic layers we will call each atomic layer 
by the medium its neighbors belong to. The atomic layer AA
will be the second from the interface into medium A. AB will be 
the last atomic layer belonging to medium A and facing the first 
atomic layer
of medium B and so on. So the four atomic layers that 
constitute the interface domain
will be labeled AA, AB, BA, and BB.
For the interfaces aligned along the (001) direction the two media
are facing each other either through its anion or cation atomic layer. 
In the alloy case, we consider a pseudobinary compound 
so that the concept of anion and cation atomic layers remain meaningful.
We will consider here only anion-anion interfaces but our results
can be extented without difficulty to other kind 
of interfaces.
We will project the interface electronic band
structure on each atomic layer and we will see how the different states that we
found for the free surface 
and for the binary/binary interface case change or disappear at the 
binary/ternary one.

It is known that the common anion interfaces
have small valence band--offset and the common cation ones have small 
conduction band--offset, both of  the order of some meV \cite{ichino,pelhos,%
duc}. In consequence, 
we will use the boundary condition that the
top of the valence bands at the interface are aligned and choose this energy
as our zero. Accordingly, the conduction band offset will be equal to the
difference in the band gaps. The actual calculation of the band 
offset is still an open theoretical question that we do not 
want to address in this work \cite{infcs2}. 
As a general remark, the FISIM states are not Bloch states and therefore
the {\bf k}--wave number is not expected to be a good quantum
number. The existence of a frontier (surface or interface) breaks
the symmetry. This does not actually mean that when the Schr\"odinger
equation is solved for differents values of {\bf k} (the Hamiltonian 
depends explicitly on it) one should get the same eigenvalue. It is 
found, theoretically \cite{prb50,prb51} as well as experimentally 
\cite{niles,gawlik}, that the solution 
does not depend on {\bf k} for the case of a surface.
In this case the boundary condition is that the wave 
function has to be zero at the surface boundary for any value
of the derivative. It is the condition for an infinite 
potential barrier. For the interface it is not so. For the 
binary/binary case we got a solution that depends on the wave 
vector, {\bf k}, but we should not call it {\it dispersion} since it 
is not the behaviour with respect to a quantum number that we 
are looking at but rather with respect to a parameter. FISIM 
states are neither Bloch states nor surface states. They 
do exist in the semi--infinite medium space but they do not 
follow the infinite--medium symmetry of the crystal. So we
have to look for a different physical reason of their {\bf k}--dependence. 

The first thing to notice is that the boundary condition is different.
For an interface, the wave function has not to be zero, is has to
be continuous together with the derivative. The boundary condition
therefore will depend on {\bf k}. This is because the Hamiltonian
describing the interaction depends on it and therefore the 
wave function that solves the Schr\"odinger equation does 
depend on it as well. For this reason its value and its 
derivative at the border will also depend on it. One does 
therefore, in general, expect a {\bf k}--dependence of the FISIM 
states eigenvalues for an interface. For a surface 
the boundary condition is always zero and on the contrary we 
do not expect a {\bf k}--dependence. 

In previous work,\cite{infcs} we have explored the behaviour 
of the FISIM states at binary/binary interfaces. These
represent a strong change at the interface. 
In this work, we explore the existence and behaviour of the 
FISIM states at interfaces that do change slowly. 
Here, ternary alloys are chosen so as to minimize stress (same 
lattice constant in both sides) and the corresponding binary/ternary 
alloy interface FISIM states are obtained. Their 
{\bf k}--dependence as expected, is minimum. So, we 
can conclude that, in general, stress is responsible
for the {\bf k}--dependence of the FISIM states. This 
is in agreement with the ideas developed above.
Therefore, FISIM states are a consequence of the 
existence of a frontier and their {\bf k}--dependence is a 
result of the stress at it. 

Furthermore, we have obtained 
from this calculation two interface 
states in the valence band range for the CdTe--based interfaces
and one interface state for the ZnSe--ones. Now we present 
the details for each interface.

In Figs. 1--4, we show the electronic band structure of the valence band for
the interfaces studied here, (001)--CdTe/CdTe$_{.85}$Se$_{.15}$,
(001)--CdTe/Zn$_{.17}$Cd$_{.83}$Te, (001)--ZnSe/ZnSe$_{.87}$Te$_{.13}$,
and  (001)--ZnSe/Zn$_{.85}$Cd$_{.15}$Se.
The dispersion relations are found from the poles
(triangles in the figures) of the real part
of the interface Green's function.
The solid--lines are a guide to the eye.
These are to be compared to the dispersion
curves found for the bulk (infinite medium) case.
The calculated eigenvalues for the FISIM states are denoted by stars, 
crosses and points; the dotted lines 
are intended only as a guide to the eye. We label the FISIM states
as $B_{Ih},\ B_{Il}$, and $B_{Is}$.
This convention follows the previous 
free (001)--surfaces study (see Refs. \cite{prb50,prb51}).
The energy eigenvalues for all the calculated states are in Tables III and IV.

\subsection{The (001)--CdTe/CdSe$_{.15}$Te$_{.85}$ interface}

Fig. 1 shows the projected electronic structure of the valence band for 
this interface per atomic layer. From the figure is evident that we 
have obtained the general pattern of the projected band structure
of the II--VI semiconductor surfaces \cite{prb50,prb51}.
As we have commented above we will consider
an anion--anion interface and we will aling the top of the 
valence band as our zero of energy. We have obtained that the heavy hole 
(hh) and light hole (lh) 
bands show more dispersion in the interface domain than in the semi--infinite
medium. They are usually low in energy about 0.7 eV an 0.4 eV, 
respectively, in all the atomic layers. The spin--orbit band shows almost the 
same dispersion that in the semi--infinite medium, see Table III. 

As is pointed previously \cite{infcs}, the FISIM states $B_{Ih}$ and 
$B_{Il}$ in the interface domain do not mix with the hh and lh bands,
as is observed in the (001)--surface case \cite{prb50,prb51}, see Fig. 1 and
Table IV. The states are lower in energy than the lh band. These 
upper FISIM states show a slight dependence on {\bf k}, but in most of 
the cases it is less than 0.3 eV. In contrast, the $B_{Is}$ 
state follows the spin--orbit band as in the semi--infinite medium. In 
general, from the Fig. 1 we appreciate that the FISIM states show
better behaviour than in the binary/binary interfaces \cite{infcs}.

Moreover, in this energy interval we have obtained some states that 
we identify with {\it interface states} ($IS_1$ and $IS_2$, the dotted
lines in Fig. 1, are a guide to the eye). The first one, at --1.3 eV,
in $\Gamma$, shows
notable dispersion and seems to disappear for {\bf k}--values
near the $X$--point. The second state, with more noticely 
dispersion, appears at --1.9 eV in $\Gamma$ and reaches the
$X-$point in --4.3 eV. However, as we do not know about experimental 
results in this system we can not give a complete comparison.
We only predict the possibility of the existence of these 
interface states.

\subsection{The (001)--CdTe/Zn$_{.17}$Cd$_{.83}$Te interface}

This system shows almost the same pattern describe above. The 
calculated valence band electronic structure is presented in 
Fig. 2. From the Table III we observe
that the hh and lh bulk bands show more dispersion that in the 
semi--infinite medium. In particular the electronic structure
projected onto the Cd--atomic layer (Fig. 2a).) shows bigger
dispersion for these bands, of about 0.8 and 0.5 eV, respectively, 
than the semi--infinite medium, see Table III. For the other 
atomic layers the projected electronic structure shows almost 
the same pattern all together: the hh and lh bands are 0.5 
and 0.3 eV below in energy with respect to the bulk values,
respectively. The spin--orbit band, however, in the interface seems 
to form a barrier of about 0.2 eV from the AA--atomic layer 
to the BB--atomic layer, see Table III. 

In general, the $B_{Ih},\
B_{Il}$ and the $B_{Is}$ FISIM states are lower in energy 
than the hh, lh, and spin--orbit bulk bands, respectively. In the same 
way that in the previous case, these FISIM states do not mix 
with the respective bulk bands at $X$, as in observed in the 
semi--infinite medium case \cite{prb50,prb51}. However,
the FISIM states shows slight dependence on {\bf k}.
As in the previous interface,
we obtain two interface states in the present system,
label $IS_1$ and $IS_2$ in Fig. 2. The $IS_1$ state, located at 
--1.3 eV in $\Gamma$, shows notable dispersion and reaches
the $X-$point between the hh and lh bulk bands. The $IS_2$ state,
with bigger dispersion than the previous one, is located
in $\Gamma$ at --1.9 eV and reaches the $X-$point at the 
same values that the spin--orbit bulk band.

\subsection{The (001)--ZnSe/ZnSe$_{.87}$Te$_{.13}$ interface}

Fig. 3 shows the calculated electronic structure of the 
valence band for this interface. Opposite to the CdTe--based
interfaces, discused above, the bulk bands and the FISIM 
calculated states for this system shows almost the same behaviour
that in the semi--infinite medium. This observation goes for 
all the calculated bands but the spin--orbit band in the $\Gamma$ point,
where we obtain, as in the previous case, a discontinuity from 
the AA--atomic layer to the BB--atomic layer. In this case 
the spin--orbit band seems to form a potential well, in $\Gamma$, of about 
0.2 eV, see Table III. On the other hand, for this interface
we obtain that the $B_{Ih},\ B_{Il}$, and $B_{Is}$ FISIM states
mix with the hh, lh, and spin--orbit bulk bands, respectively, as is 
observed in the semi--infinite medium \cite{prb50,prb51}. In this 
sense this interface shows better behaviour than the other ones \cite{infcs}.
Although, as previously, we have obtained an interface state for this 
system, $IS_1$. This interface state appears in $\Gamma$ at --1.7 eV, shows
noticely dispersion and seems to disappear for {\bf k}--values near
the $X-$point. However, the state appears notoriously in all the 
calculated atomic layers.

\subsection{The (001)--ZnSe/Zn$_{.85}$Cd$_{.15}$Se interface}

Finally, in the Fig. 4 we show our electronic structure calculated
for this system. In the same way that the previous case, we obtain
that all the calculated states, per atomic layer, for this interface
are similar with the semi--infinite medium, see Tables III and IV.
In addition to these states, we have an interface state, $IS_1$, located
in $\Gamma$ at --1.7 eV and showing notable dispersion. The state 
do not appear for all the interval between $\Gamma-X$, it seems to 
disappear for {\bf k}--values near the $X-$point, as we have 
commented previously for the other interfaces.

\section{Conclusions}

In conclusion, we 
have calculated the electronic structure of the valence band of the 
II--VI binary/ternary alloy interfaces. We have used the tight--binding
method and the surface Green's function matching method to obtain 
the electronic structure projected onto each atomic layer that 
constitutes the 
interface domain. For the ternary alloys we have used our 
tight--binding Hamiltonians described in previous work that give 
good account for the changes of the band gap with composition as obtained
experimentally. Our parametrization includes an \ebp\ for the ``$s$'' on--site
tight--binding parameter of the substituted ion and we 
use the known virtual crystal approximation for the rest of 
them. The systems were chosen here so that stress can be 
ignored for the particular value of the compositional variable.
The calculated valence band electronic structure of these interfaces 
show bulk bands with similar dispersion as for the semi--infinite
medium (a system with a surface). 
The FISIM states observed in the (001)--oriented 
surfaces and binary/binary interfaces appear also in this case and 
show an intermediately strong {\bf k}--dependence as compare to the 
previous ones. 
In the interface domain the calculated states,
both the bulk bands and the FISIM states, have a composition that is 
a combination
of the corresponding states of the two media forming the interface.

It is interesting to note further that we have
obtained for the binary/ternary alloy case 
two interface states for the CdTe--based
heterostructures and one interface state for the ZnSe--ones that do not 
show for the binary/binary interfaces at least in the energy interval
that we have considered. We will consider the binary/quaternary 
and the ternary/quaternary alloy interfaces in future work.


\begin{figure}
\caption{Electronic structure of the valence band, per atomic layer,
of the \hfill\break (001)--CdTe/CdSe$_{.15}$Te$_{.85}$ interface. The dispersion
relations are obtained from the poles (triangles) of the real part
of the interface Green's function. The solid lines are a guide to
the eye. $B_{Ih},\ B_{Il}$ and $B_{Is}$ are the calculated FISIM states
(stars, crosses and points, the dotted lines, intended to 
show the dispersion, are a guide to the eye). We show the 
interface states $IS_1$ and $IS_2$.}
\end{figure}

\begin{figure}
\caption{Electronic structure of the valence band of the 
(001)--CdTe/Zn$_{.17}$Cd$_{.83}$Te interface. See Fig. 1 for
details.}
\end{figure}

\begin{figure}
\caption{Electronic structure of the valence band of the 
(001)--ZnSe/ZnSe$_{.87}$Te$_{.13}$ interface. See Fig. 1 for
details.}
\end{figure}

\begin{figure}
\caption{Electronic structure of the valence band of the 
(001)--ZnSe/Zn$_{.15}$Cd$_{.85}$Se interface. See Fig. 1 for
details.}
\end{figure}

\newpage

\begin{table}
\caption{
Empirical bowing parameter for the ternary alloys used in this work.
Taken from Ref. [18].
}
\begin{tabular}{||c|cc||}
               Compound             &       b$_a$   &      b$_c$     \\
\hline
        ZnSe$_{1-x}$Te$_x$          &    --6.964    &      $-$       \\
        CdSe$_{1-x}$Te$_x$          &    --0.195    &      $-$       \\
\hline
        Zn$_{1-x}$Cd$_x$Se          &     $-$       &     0.037      \\
        Zn$_{1-x}$Cd$_x$Te          &     $-$       &     0.020      \\
\end{tabular}
\label{ebp}
\end{table}

\mediumtext
\begin{table}
\caption{Lattice parameter ratio for the selected composition 
and the induced stress for the 
binary/ternary alloy interfaces studied in the present work.}
\begin{center}
\begin{tabular}{||c|c|c||}
Interface    & Lattice parameter ratio & Induced stress \% \\
\hline
CdTe/CdSe$_{.15}$Te$_{.85}$& 6.481/6.4175  &  1.    \\
CdTe/Zn$_{.17}$Cd$_{.83}$Te& 6.481/6.4171  &  1.    \\
ZnSe/ZnSe$_{.87}$Te$_{.13}$& 6.052/5.7239  &  1.    \\
ZnSe/Zn$_{.85}$Cd$_{.15}$Se& 5.052/5.7273  &  1.    \\
\end{tabular}
\end{center}
\label{red}
\end{table}

\begin{table}[!t]
\caption{
Energy--eigenvalues for the heavy hole (hh), light hole (lh), 
and spin--orbit 
bands at $\Gamma$ and $X$ high--symmetry points as obtained for the
interface domain for (001)--CdTe/CdSe$_{.15}$Te$_{.85}$, 
(001)--CdTe/Zn$_{.17}$Cd$_{.83}$Te,
(001)ZnSe/ZnSe$_{.87}$Te$_{.13}$, and
(001)--ZnSe/Zn$_{.85}$Cd$_{.15}$Se. The energies are in eV.
}
\begin{center}
\begin{tabular}{||c|c|c|ccc||}
  &    & $\Gamma-$point &  & $X-$point  &  \\
\hline
\hline
System & Atomic layer& $E_{so}$ & $E_{hh}$ & $E_{lh}$ & $E_{so}$ \\
\hline
                &Cd   & --0.8 & --2.4 & --2.6 & --4.2  \\
                &Te   & --0.8 & --2.3 & --2.6 & --4.2  \\
CdTe / CdSe$_{.15}$Te$_{.85}$   &Se$_{.15}$Te$_{.85}$& --0.8 & --2.4 & --2.6 & --4.2 \\
                &Cd   & --0.8 & --2.3 & --2.6 & --4.2  \\
\hline
                &Cd   & --1.0 & --2.5 & --2.7 & --4.3  \\
                &Te   & --0.8 & --2.2 & --2.5 & --4.1  \\
CdTe / Zn$_{.17}$Cd$_{.83}$Te&Te   & --0.8 & --2.2 & --2.5 & --4.1  \\
                &Zn$_{.17}$Cd$_{.83}$& --1.0 & --2.3 & --2.5 & --4.3  \\
\hline
                &Zn   & --0.4 & --1.9 & --2.4 & --5.0  \\
                &Se   & --0.6 & --1.9 & --2.4 & --5.0  \\
ZnSe / ZnSe$_{.87}$Te$_{.13}$&Se$_{.87}$Te$_{.13}$& --0.6 & --1.9 & --2.4 & --5.0  \\
                &Zn   & --0.4 & --1.9 & --2.4 & --5.0  \\
\hline
                &Zn   & --0.4 & --2.0 & --2.2 & --5.0  \\
                &Se   & --0.4 & --2.0 & --2.2 & --5.2  \\
ZnSe / Zn$_{.85}$Cd$_{.15}$Se&Se   & --0.4 & --2.0 & --2.2 & --4.9  \\
                &Zn$_{.85}$Cd$_{.15}$& --0.4 & --2.0 & --2.2 & --5.0  \\
\end{tabular}
\end{center}
\label{tabla2}
\end{table}

\begin{table}[!t]
\caption{
Energy--eigenvalues for the FISIM states (see text), 
$B_{Ih},\ B_{Il}$, and $B_{Is}$,
at the interface dominion of
(001)--CdTe/CdSe$_{.15}$Te$_{.85}$,
(001)--CdTe/Zn$_{.17}$Cd$_{.83}$Te,
(001)ZnSe/ZnSe$_{.87}$Te$_{.13}$, and
(001)--ZnSe/Zn$_{.85}$Cd$_{.15}$Se.
}
\begin{center}
\begin{tabular}{||c|c|ccc|ccc||}
      &  & $\Gamma-$point  &      &    & $X-$point    &    &     \\
\hline
\hline
System & Atomic layer& $B_{Ih}$  & $B_{Il}$ & $B_{Is}$ & $B_{Ih}$ & $B_{Il}$
                                                            & $B_{Is}$ \\
\hline
          & Cd      & --2.4 & --2.5  & --4.5   & --2.6 & --3.1 & --4.6  \\
          & Te      & --2.4 & --2.5  & --4.5   & --2.6 & --3.1 & --4.6  \\
CdTe/CdSe$_{.15}$Te$_{.85}$&Se$_{.15}$Te$_{.85}$& --2.4 & --2.5  & --4.5   & --2.4 & --3.0 & --4.6  \\
          &   Cd    & --2.4 & --2.5  & --4.5   & --2.6 & --3.0 & --4.6  \\
\hline
          & Cd      & --2.5 & --2.9  & --4.6   & --2.7 &  --3.2 & --4.7 \\
          & Te      & --2.5 & --2.7  & --4.5   & --2.5 &  --3.0 & --4.5 \\
CdTe/Zn$_{.17}$Cd$_{.83}$Te&Te      & --2.5 & --2.9  & --4.5   & --2.5 &  --2.9 & --4.7 \\
          & Zn$_{.17}$Cd$_{.83}$& --2.5 & --2.9  & --4.7   & --2.7 &  --3.2 & --4.8 \\
\hline
          & Zn      & --2.0 & --2.2  & --5.1   & --1.9 & --2.4 & --5.2 \\
          & Se      & --2.0 & --2.2  & --5.2   & --1.9 & --2.4 & --5.2 \\
ZnSe/ZnSe$_{.87}$Te$_{.13}$&Se$_{.87}$Te$_{.13}$& --2.0 & --2.2  & --5.2   & --1.9 & --2.4 & --5.2 \\
          & Zn      & --2.0 & --2.2  & --5.1   & --1.9 & --2.4 & --5.2 \\
\hline
          & Zn      & --2.0 & --2.2  & --5.1   & --2.2 & --2.4 & --5.2 \\
          & Se      & --2.0 & --2.2  & --5.1   & --2.2 & --2.4 & --5.2 \\
ZnSe/Zn$_{.85}$Cd$_{.15}$Se&Se     & --2.0 & --2.2  & --5.1   & --2.2 & --2.4 & --5.1 \\
          & Zn$_{.85}$Cd$_{.15}$& --2.0 & --2.2  & --5.1   & --2.2 & --2.4 & --5.2 \\
\end{tabular}
\end{center}
\label{tabla3}
\end{table}

\end{document}